\documentclass[%
reprint,
amsmath,amssymb,
aps,
pra,
notitlepage
]{revtex4-1}
\usepackage[english]{babel}
\usepackage[utf8]{inputenc}
\usepackage{amsmath}
\usepackage{physics}
\usepackage{graphicx}
\graphicspath{{figures/}}
\usepackage[colorinlistoftodos]{todonotes}
\usepackage{lipsum}
\usepackage{multirow}
\usepackage{calc}
\usepackage[bottom]{footmisc}
\usepackage[]{hyperref}
\usepackage{booktabs}
\usepackage[normalem]{ulem}
\usepackage{float}
\usepackage{lineno}

\addto\captionsenglish{}

\DeclareUnicodeCharacter{2212}{-}

\begin{document}

\title{Engineering AlGaAs-on-insulator towards quantum optical applications}
\author{Marlon Placke$^{1,2,\star}$ and Sven Ramelow$^1$}
\affiliation{$^{1}$ Humboldt-Universit\"{a}t zu Berlin, Newtonstra\ss e 15, 12489 Berlin \\
$^{2}$Ferdinand-Braun-Institut f\"{u}r H\"{o}chstfrequenztechnik, Gustav-Kirchoff-Stra\ss e 4, 12489, Berlin \\
$^{\star}$ Corresponding author: mplacke@physik.hu-berlin.de}
\begin{abstract}
Aluminum gallium arsenide has highly desirable properties for integrated parametric optical interactions: large material nonlinearities, maturely established nanoscopic structuring through epitaxial growth and lithography, and a large band gap for broadband low-loss operation. However, its full potential for record-strength nonlinear interactions is only released when the semiconductor is embedded within a dielectric cladding to produce highly confining waveguides. From simulations of such, we present second and third order pair generation that could improve upon state-of-the-art quantum optical sources and make novel regimes of strong parametric photon-photon nonlinearities accessible.
\end{abstract}
\maketitle
Entangled photons lie at the very heart of photonic quantum technology. The favored workhorse technique for their generation has long been spontaneous parametric down conversion (SPDC) from second order nonlinear crystals \cite{PhysRevA.63.062301}, due to their comparatively easy room-temperature operation, high spectral brightness, and ability to produce various forms of entanglement by setup design. For an efficient interaction, one must, however, carefully match the phase velocities of the pump and signal/idler fields (phase matching) to compensate the dispersion of the nonlinear medium. Quasi-phase-matching (QPM) offers an elegant and efficient solution, but also restricts the type of crystal to be used for efficient photon pair sources to materials such as LiNbO$_3$, KTP or SLT, even when other materials would, in principle, offer significantly larger nonlinearities \cite{doi:10.1063/1.336070} or other beneficial functionalities.\\
Amongst the latter reside direct band gap semiconductors with maturely established fabrication techniques that, hence, could enable highly efficient nonlinear devices with co-integrated pump lasers. 
For possessing these features while also being compatible with telecom frequencies, aluminum gallium arsenide has been heralded as the "silicon of nonlinear optics" \cite{doi:10.1142/S0218199194000213}. However, the low confinement that is commonly achieved in fabrication friendly (i.e. monolithic) phase matching approaches such as Bragg reflection waveguides (BRW) limits their performance \cite{Kang:16, PhysRevLett.108.153605}. Such structures typically feature length-normalized efficiencies on the order of $\tilde{\eta} \sim 10^{-7}$ cm$^{-3/2}$ which state-of-the art periodically-poled lithium niobate (ppLN) based sources outperform by an order of magnitude \cite{ Chen:19, PhysRevLett.124.163603}. Here, for direct comparison, the choice of dimensions normalizes the absolute efficiency $\eta$ for the $L^{3/2}$ length-dependence that is characteristic for degenerate down conversion\footnote{A comparison of absolute efficiencies might, nevertheless, better account for technical limitations that impose boundaries on accessible device lengths such as propagation losses or film thickness variations of the deposited materials.}.
\\ In contrast to this, fully surrounding the semiconductor core with dielectric cladding enables large index contrasts and highly tunable dispersion properties. In combination with sub-micrometer mode confinement, the effective nonlinearities of these structures increase drastically such that nonlinear processes can already become significant at very low levels of pump (seed) power. Such AlGaAs-on-insulator waveguides have recently been realized through flip-bonding the dielectric-capped AlGaAs layer onto a carrier substrate and enclosing the waveguide (WG) core in the same dielectric after its lithographic structuring. This has born impressive results in classical nonlinear optical interactions e.g. broadband Kerr frequency comb generation \cite{Pu:16} as well as record efficiency on-chip second harmonic generation  (SHG) - albeit for telecom-wavelength incompatible GaAs  \cite{chang2018heterogeneously}.
Furthermore, realizing parametric photon-photon nonlinearities has, in the past decade, become a distinct research effort of nonlinear quantum optics after numerous proposals have identified feasible strategies to lower seed power requirements towards single photon power levels  \cite{Langford2011, PhysRevLett.110.223901, Guerreiro2013, Venkataraman2013, PhysRevA.90.043808}. The anticipated transformative advancement for scalable photonic quantum information processing that such interfaces would manifest continues to motivate an expanded quest of tailoring nonlinear platforms towards such functionality \cite{doi:10.1063/1.4913743, PhysRevLett.116.023601, PhysRevLett.118.123601, PhysRevLett.120.160502, PhysRevLett.122.153906, PhysRevApplied.13.044013, PhysRevLett.124.160501}.
Our focus lies with the potential of AlGaAs-on-insulator to advance parametric nonlinearities to unprecedented regimes. We present sources of SPDC and four-wave mixing (FWM) induced pair generation at (application friendly) telecom wavelengths and optimize their geometry for highest efficiency. Our one-dimensional SPDC optimization demonstrates competitive efficiencies for the promising AlGaAs-on-insulator platform comparable to the mature performance of state-of-the-art ppLN WGs. Finally, we demonstrate the flexibility of a novel two-dimensional approach to dispersion engineering that predicts the improvement of resonator based FWM efficiencies by orders of magnitude.\\
\begin{figure}[htbp]
\centering
\includegraphics[width=0.9\linewidth]{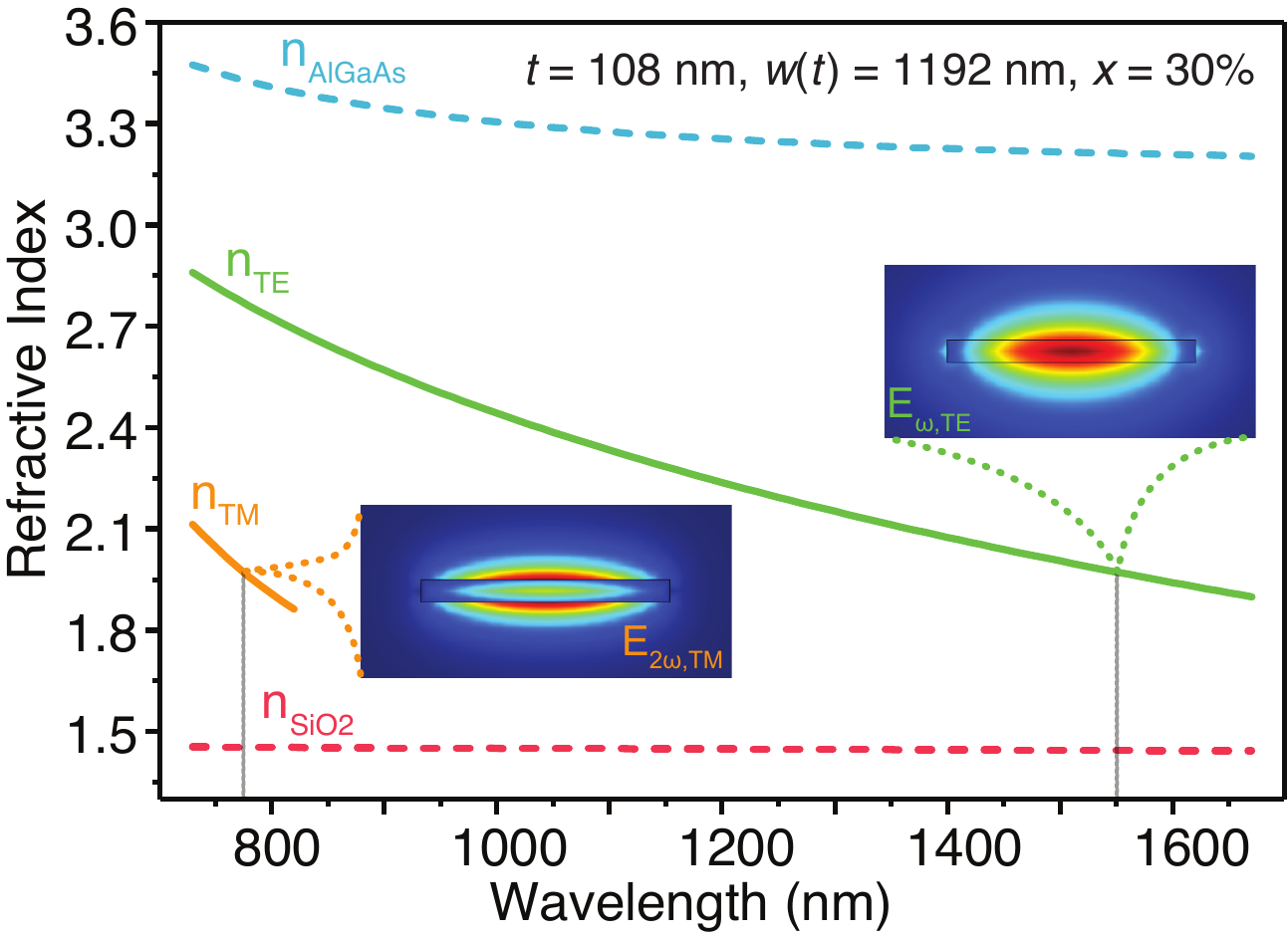}
\caption{\small{Simulated dispersion of the horizontally and vertically polarized fundamental modes in an Al$_{0.3}$GaAs/SiO$_2$ WG. The strongly form-birefringent core geometry ($108 \times 1192$ nm$^2$) features matching indices at the (biphoton) wavelength of 1551.4 nm and its corresponding second harmonic. The insets show the electric field distributions of the main polarization components of both modes near the phase matching point.}}
\label{fig:Fig1}
\end{figure}
For the SPDC phase matching, we make use of the strong form-birefringence of thin film WG core geometries similar to \cite{chang2018heterogeneously}. In our simulations, a rectangular Al$_{x}$Ga$_{1-x}$As core is embedded in Si$O_2$ cladding. For a given thickness of the core, its width is chosen such that the larger evanescent amplitude in the cladding of the short wavelength (TM$_{00}$-like) pump mode averages to the same effective index as the (TE$_{00}$-like) signal/idler mode at 1550$\pm 5$ nm. The evanescent amplitude is especially pronounced in thin film geometries, where the fundamental TM mode exhibits large discontinuities at the film interfaces (see Fig. \ref{fig:Fig1} insets). A finite element solver (COMSOL) is used to simulate the dispersion of the two fundamental modes\footnote{We refined the computational mesh to achieve a numerical accuracy of <10$^{-4}$ for the simulated refractive index. This numerical accuracy corresponds to a temperature variance of < 0.5 K in the empirical model that was used for the material refractive indices that feed the simulations [22].} and their respective field distributions $\vec{E}_{m}(\zeta, \xi)$. The latter quantify the effective nonlinear overlap $\Gamma_d$ as an average of the core and cladding material susceptibilities $d_{\textrm{mat}}(\zeta, \xi)$ weighted by a three-field mode overlap:
\begin{align}
 \Gamma_d = \frac{d_{\textrm{eff}}}{\sqrt{A_{\textrm{eff}}}} \approx \frac{\iint_{\textrm{core}} d_{\textrm{\tiny{AlGaAs}}} E_{2 \omega, \textrm{TM}}^*(\zeta, \xi) E_{\omega, \textrm{TE}}^2(\zeta, \xi) \textrm{d}\zeta \textrm{d}\xi}{\left[ \iint  \left| \vec{E}_{2 \omega} \right|^2 d\zeta d\xi \right]^{1/2}  \iint \left| \vec{E}_{\omega}\right|^2 \textrm{d}\zeta \textrm{d}\xi } \label{eq:refname2}    \ .
\end{align}
Due to the off-diagonal tensorial nature of the nonlinearity of AlGaAs ($d_{ijk}\neq 0 \leftrightarrow i\neq j \neq k$), only the main polarization components of the two modes make a significant contribution to $\Gamma_d$
and the WGs must propagate in <110>-direction for a maximum projection of the TE mode on the crystallographic $ij$-axes.
Because the linear power scaling of SPDC intersects with the quadratic scaling of second harmonic generation (SHG) when either interaction is pumped with the equivalent power of one signal/idler photon per coherence time ($\tau = \Delta \nu^{-1}$), the SPDC efficiency follows from rearranging the equivalent expression for SHG \cite{Helt:12}, accordingly 
\begin{equation}
    \eta_{\textrm{spdc}} = \frac{P_{\textrm{shg}}(P_{\textrm{p}} = h \nu_0 \Delta \nu)}{P_{\textrm{p}}} = \frac{8 \pi^2 h L^2}{\lambda^3_0 \epsilon_{\textrm{0}} n^3_{\textrm{eff}}} \Gamma^2_d \Delta \nu= \tilde{\eta}_{\textrm{spdc}} L^{3/2} \quad .
\end{equation}

\begin{figure}[htbp]
\centering
\includegraphics[width=\linewidth]{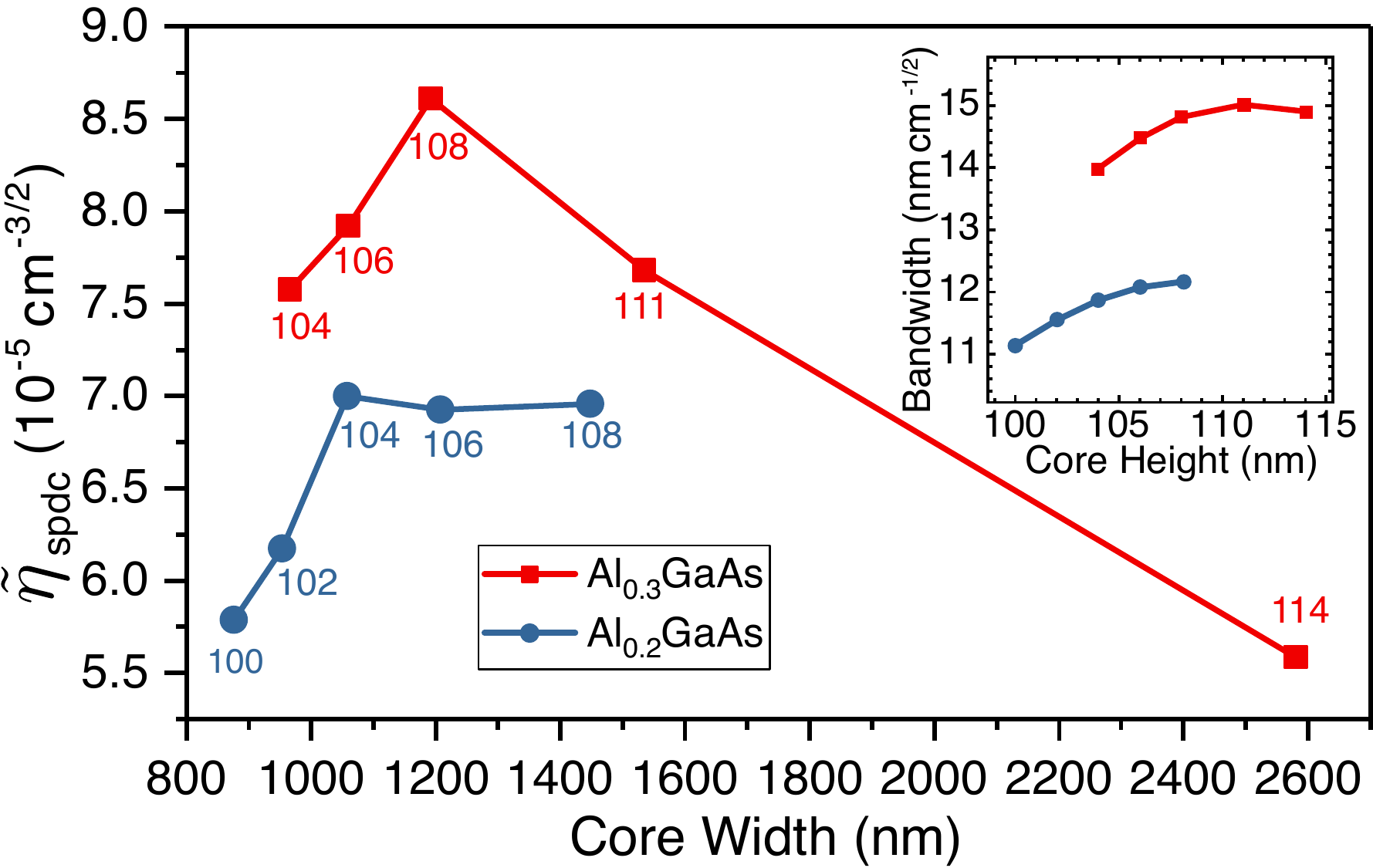}
\caption{\small{Length-normalized SPDC efficiency as a function of core geometry: The numbers near each data point indicate the corresponding core thickness in nm. Owing to larger signal/idler bandwidths (inset), the Al$_{0.3}$GaAs WGs feature higher efficiencies despite offering the smaller material nonlinearity compared to Al$_{0.2}$GaAs. }} 
\label{fig:Mgraph}
\end{figure}
Fig. \ref{fig:Fig1} shows the simulated dispersion of the TE$_{00}$-like and TM$_{00}$-like mode in a 108 nm thick and 1192 nm wide Al$_{0.3}$GaAs core in SiO$_2$ cladding and their respective modal field distributions near the phase matching point ($\lambda_0 = $ 1551.4 nm).
To find the optimum SPDC efficiency, we compare performance figures of five phase matched geometries with core thicknesses from 100 to 108 nm and corresponding widths from 875 to 1450 nm and an aluminum fraction of 20\% with another set of five phase matched geometries with film thicknesses from 104 to 114 nm and corresponding widths from 966 to 2580 nm and 30\% aluminum, respectively\footnote{The efficiencies were calculated with material nonlinearities of $d = 114$ pm/V and $d = 110$ pm/V for AlGaAs with an aluminum fraction of twenty and thirty percent, respectively [24,25].}.\\
Fig. \ref{fig:Mgraph} shows both efficiency distributions to feature the peak value at the central core geometry. Because the smaller material nonlinearity is overcompensated by an increased phase matching bandwidth (see Fig. \ref{fig:Mgraph} inset), the largest efficiency of $\tilde{\eta}_{\textrm{spdc}} \approx 8.6 \cdot 10^{-5}$cm$^{-3/2}$ is predicted for the higher aluminum concentration. Nevertheless, as both the material nonlinearity decreases nonlinearly and the surface oxidization becomes more significant with further increasing Al concentration, we limited the range of interest to $x \leq 30$ \%,  while the lower boundary (of $x =$ 20 \%) ensures transparency at the pump wavelength.
\begin{figure}[htbp]
\centering
\includegraphics[width=\linewidth]{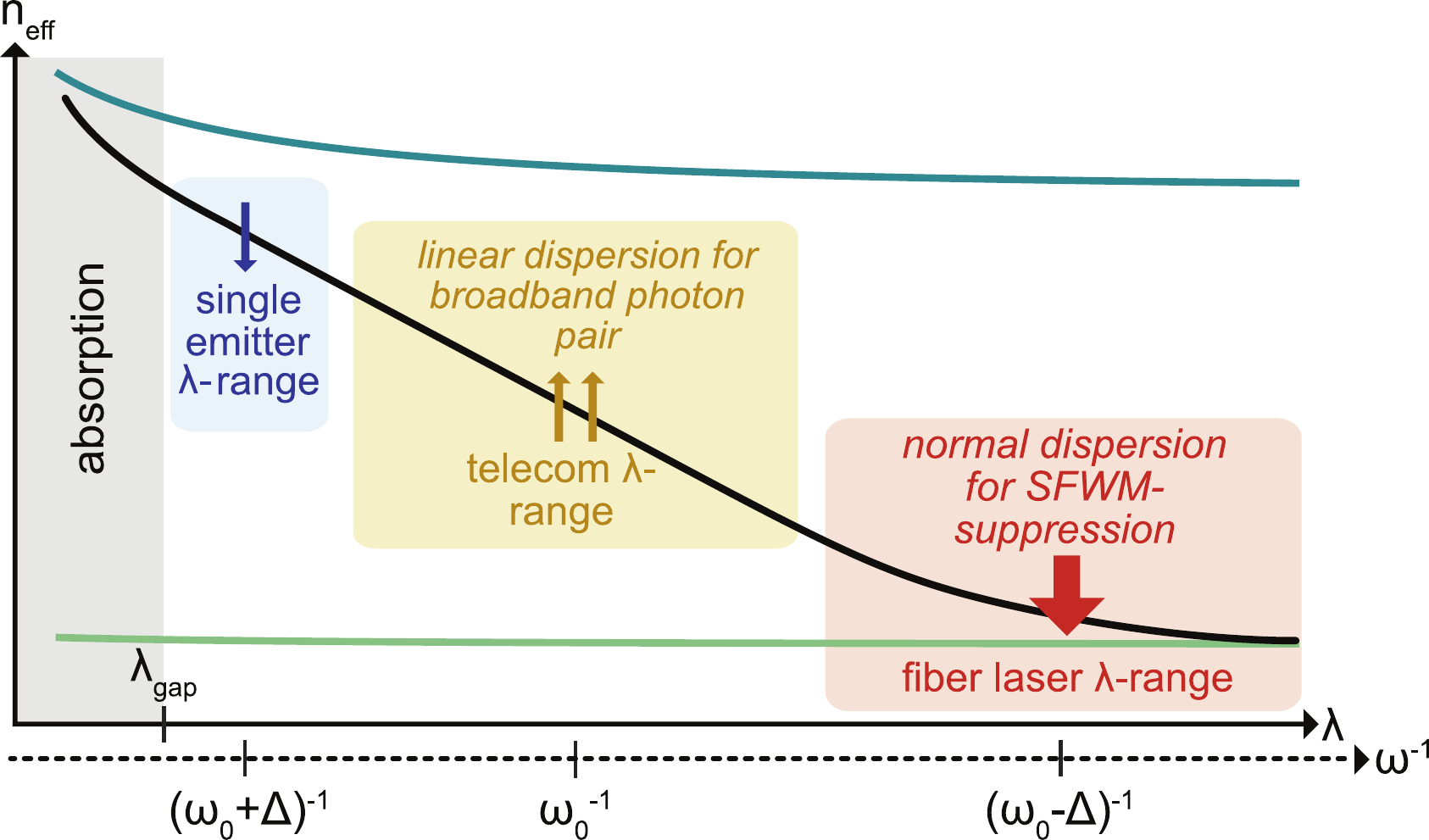}
\caption{\small{Scheme for coherent photon conversion: A strong red-detuned pump is used to scale the coherent coupling of signal/idler photons at center frequency with an equally blue-detuned single photon pump.}} 
\label{fig:chi3_sketch}
\end{figure}
\\By comparison with experimentally realized integrated sources, the proposed structures enable broadband SPDC that is competitive with state-of-the-art ppLN-based sources. For the latter, absolute efficiencies (per pump photon) of up to $\eta_{\textrm{spdc}}$ = 6 $\cdot 10^{-7}$ were demonstrated in 4 mm long WGs with a signal/idler bandwidth of 35 nm \cite{Chen:19}, while an extrapolation from simulated SHG values suggests a theoretical performance of $\eta_{\textrm{spdc}}$ = 1.4 $\cdot 10^{-6}$ \cite{Wang:18}. In contrast, the AlGaAs WG with efficiency-optimized geometry achieves the same bandwidth for a length of 1.7 mm, for which an absolute efficiency of up to $\eta \approx 6 \cdot 10^{-6}$ and a corresponding spectral brightness of $\phi_{\nu} \approx 5.5 \cdot 10^{6}$ (s mW GHz)$^{-1}$ are predicted.
These performance figures enable highly efficient quantum key distribution (QKD) sources that utilize telecom infrastructure as the following estimate illustrates:  Assuming a temporal resolution in detection of $\Delta t = 200$ ps, and trading-off double emission events against brightness, only 0.8 mW of pump power is required to produce 0.05 photon pairs per time bin per (55 GHz broad) dense wavelength division multiplexing (DWDM) channel. As such power levels can be provided by simple laser diode structures, the on-chip integration of the pump laser and the SPDC source seems feasible from the standpoint of the optical power requirements.
\\Finally, the short length required for broadband phase matching makes the AlGaAs WGs very tolerant to the most common loss factors, namely film thickness variation and scattering loss (of the short wavelength pump) from surface roughness. We verify that a 1.7 mm short WG is very tolerant to pump propagation losses, suffering only a 3 dB (10 dB) reduction in total efficiency for -30 dB/cm (-110 dB/cm) of loss \footnote{Based on the comparison of $\eta(\alpha) \propto \int _0 ^L z \exp{(-\alpha z)} \textrm{d} z$ with the (lossless) efficiency.}. Furthermore, the thickness variation of flip-bonded epitaxial films has been reported to become relevant only for several millimeter long WGs \cite{Stanton:20}.
\\
Similar to the second order case, (Al)GaAs provides a large third order material nonlinearity to begin with and the effective nonlinearity is even more efficiently enhanced by mode confinement ($\gamma = 2 \pi n_2 /(\lambda A_{\textrm{eff}})$). 
By virtue of this scaling, FWM schemes can reach efficiencies that allow for significant nonlinear optical interactions at single photon power levels. To also illustrate the potential of the AlGaAs-on-insulator platform for fundamental research applications, we expanded the simulations to yield the performance figures of so-called coherent photon conversion (CPC) schemes \cite{Langford2011}.
In our CPC implementation, a strong, red-detuned pump pulse is used to scale the coherent coupling of a photon pair at center frequency with an equally blue-detuned single photon (Fock) pump mode, see Fig. \ref{fig:chi3_sketch}.\\
Because of the high pump power, a phase shift from self- and cross-phase modulation must be accounted for. The overall phase mismatch function becomes \cite{PhysRevA.90.043808}
\begin{equation}
 K(\omega) = \beta_2 (\Omega^2 - \Delta^2) + \frac{\beta_4}{12} (\Omega^4 - \Delta^4) + \gamma P \ \ ,
\end{equation}
where $\Omega = \omega - \omega_{\textrm{0}}$ is the frequency variable of the Taylor expansion of the mode propagation constant ($\beta(\omega)$) about $\omega_0$ and $\Delta$ is the (fixed) detuning of either pump. Phase matching is achieved by adjusting the peak power of the strong pump such that the phase mismatch vanishes at center frequency ($K(\omega_0) = 0 \rightarrow P_0$) and the corresponding signal/idler bandwidth $\Delta \omega_\textrm{si}$ follows from integration of $\textrm{sinc}^2(K(\omega, P = P_0) L/2)$.
The efficiency of converting the Fock pump (at $\omega = \omega_\textrm{F}$) into the signal/idler pair is given by
\begin{equation}
    \eta_{1\rightarrow 2} = \frac{\gamma^2 P_0 \hbar \omega_{\textrm{F}} L^2 \Delta\omega_{\textrm{si}}}{\pi} = \tilde{\eta}L^{3/2}P_0  \ \  .
\end{equation}
The flexibility of this nonlinear phase matching approach allows independent choice of core width and thickness.
Accordingly, we scan over a two dimensional grid of WG cross sections to identify geometries with desirable dispersion properties - namely, a linear behavior of $n_{\textrm{eff}}(\lambda)$ around $\lambda_0$ enables a large phase matching bandwidth and a large normal dispersion around $\omega_p = \omega_0-\Delta$ suppresses spontaneous four wave mixing, see Fig. \ref{fig:chi3_sketch} for illustration.
To further bridge the gap between the fundamental physics interest of testing the material platform to host single photon-level nonlinear interactions and the usefulness and practicality of such, we fix the wavelengths of the two pumps to 960 nm and 2060 nm such that the signal/idler photon pair is centered in the telecom O-band (at 1310 nm). The choice of pump wavelengths accounts for the availability of single photon sources on the high energy side \cite{doi:10.1063/5.0010193} and strong pump lasers on the low energy side (Thulium doped fiber lasers). Given these two pump wavelength regions, (aluminum-free) gallium arsenide is chosen as the core material because it offers both the largest material nonlinearity ($n_2 = 3.3 \cdot 10^{-17}$ m$^2$W$^{-1}$ \cite{Boyd:2013wk}) and the largest index contrast to the SiO2 cladding while remaining free of one- and two-photon absorption, respectively.\\
The results of these scans are given in Fig. \ref{fig:eta}: A peak efficiency of 18 \% cm$^{-3/2}$ is found for the core geometry of 250$\times$400 nm$^2$. For this particular WG geometry, broadband phase matching ($\Delta \lambda_{\textrm{si}} \cdot L^{-1/2}  \approx 129 \ \textrm{nm}\cdot \textrm{cm}^{-1/2}$) coincides with an increased pump power requirement of $P_0 \approx 948$ W for the nonlinear phase shift that is, nevertheless, within reach of the peak powers that pulsed fiber lasers provide. Moreover, the discontinuities in the calculated bandwidth maxima indicate that the current grid undersamples the parameter space, such that further refinement of the simulation grid might reveal even higher numerical efficiencies.
\begin{figure*}[!htbp]
\centering
\includegraphics[width=15cm]{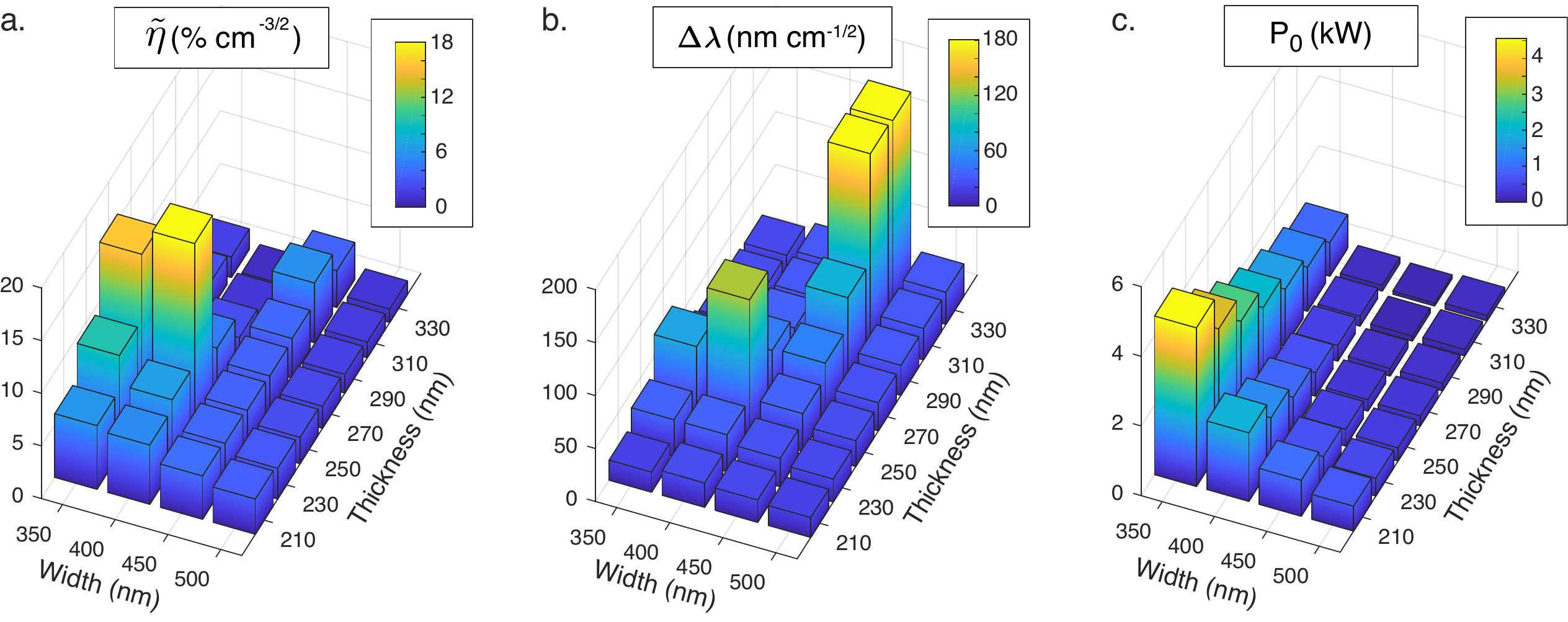}
\caption{\small{ {\bf a.} Length-normalized CPC efficiency: a peak value of 18 \% cm$^{-3/2}$ was found for the $250\times 400$ nm$^2$ GaAs core in 3 $\mu$ m of SiO$_2$ cladding.{\bf b.} and {\bf c.} Inspection of the corresponding parameters of the nonlinear phase matching shows the coincidence of a large signal/idler bandwidth with an increased phase matching power requirement at the geometry of peak efficiency.}} 
\label{fig:eta}
\end{figure*}
\\Concerning the more practical aspects of the implementation, the large nonlinear contribution ($\gamma P_0$ $\gg 2 \pi$) makes the phase matching very intolerant to propagation losses of simple WG structures, especially when approaching the regime of near-deterministic conversion ($\eta_{1\rightarrow 2}\rightarrow 1)$ demands interaction lengths of several centimeters.\\
Alternatively, cavity-based approaches e.g. ring resonator cavities could circumvent the loss-induced scaling limitation. A back of the envelope calculation for the parameters of the $250\times 400$ nm$^2$ geometry substituted into\cite{PhysRevLett.122.153906}
\begin{equation}
\eta_{1\leftrightarrow2}=\frac{256 R^{2} \gamma^{2}\left(\Delta \lambda_{F S R}\right)^{4}}{\pi^{2} \lambda^{4}} P_{P}^{i n} \hbar \omega^2 Q^{3}\left(\frac{Q}{Q^{e x t}}\right)^{4}
\end{equation}
yields a conversion efficiency of $\eta_{1\leftrightarrow2} = 2.3$ \% per resonance for a ring resonator of radius R = 50 $\mu$ m with a (1dB/cm) loss-limited loaded quality factor of $Q  = \frac{\pi n_{\textrm{eff}}}{\alpha \lambda} \approx 260 $k at critical coupling. With the phase matching bandwidth of the $250\times 400$ nm$^2$ geometry spanning over seventeen free spectral ranges in the resonator (with a coherence length of 35 cm), pair generation of near-unity efficiency ($\eta_{1\rightarrow2} = \eta_{1\leftrightarrow2} \cdot \Delta \lambda_{\textrm{si}}/\Delta \lambda_{\textrm{FSR}}   \gg 10\%$) would be approached so closely that the simplified interaction model that employs a classical second pump (``seed'') field and neglects its depletion breaks down. Compared to experimentally demonstrated state-of-the-art efficiencies reported for degenerate FWM in Si$_3$N$_4$ microring resonators, the simulated $\eta_{1\leftrightarrow2}$ values predict a potential improvement by more than three orders of magnitude \cite{Lu2019, PhysRevLett.122.153906}. Studying this regime of strong photon-photon nonlinearities both experimentally and theoretically could elucidate novel quantum optical phenomena.
\\We have demonstrated flexible phase matching schemes for second and third order nonlinear interactions at telecom wavelengths on the (Al)GaAs-on-insulator material platform. In both cases, we scanned over geometric variations that were large enough to identify peak efficiency WG models. With a required pump power on the order of 1 mW for key-rate optimized photon pair generation throughout the entire C-band, the proposed structures challenge mature performance figures of ppLN based sources while opening new perspectives for scalability and integration. Finally, the simulated CPC efficiencies are evidence that parametric nonlinear interactions can be tailored to already become significant at single photon (pump) power level. As a host to this virtue, the presented WG models could uncover novel quantum optical phenomena of strong photon-photon nonlinearities.
\section*{Funding Information}
MP acknowledges the support of the Ferdinand-Braun-Institut. SR acknowledges funding from the Deutsche Forschungsgemeinschaft (DFG) (RA 2842/1-1).
\section*{Disclosures}
The authors declare no conflicts of interest.


%

\end{document}